\documentclass[aps,prb,twocolumn,superscriptaddress,floatfix]{revtex4-2}

\usepackage{amsmath,amsfonts,amssymb}
\usepackage{natbib}
\usepackage{graphicx,color}

\usepackage{bm}

\newcommand{\mni}{\mathrm{i}}

\begin{document}

\title{Electromagnetics of deeply subwavelength metamaterial particles}

\date{}

\author{Aleksander O. Makarenko}
\affiliation{Department of Physics, ITMO University, St.\,Petersburg, Russia}

\author{Maxim A. Yurkin}
\affiliation{Universit\'{e} Rouen Normandie, INSA Rouen Normandie, CNRS, CORIA UMR 6614, Rouen, France}

\author{Alexey A. Shcherbakov}
\affiliation{Department of Physics, ITMO University, St.\,Petersburg, Russia}

\author{Mikhail Lapine}
\affiliation{School of Mathematical and Physical Sciences, University of Technology Sydney, NSW, Australia}
\affiliation{Qingdao Innovation and Development Centre of Harbin Engineering University, Huangdao, China}

%%%%%%%%%%%%%%%%%%%%%%%%%%%%%%%%%%%%%%%%%%%%%%%%%%

\begin{abstract}
This article discusses electromagnetic properties of volumetric metamaterial samples with essentially discrete structure, that is, assembled as a periodic array of electromagnetic resonators. 
We develop an efficient numerical procedure for calculating quasi-static electromagnetic response precisely to analyse samples containing several million meta-atoms.
We demonstrate that, contrary to a common belief, even million-``atoms'' samples with sharp edges are still quite different from uniform (``homogenised'') materials, and their properties are critically sensitive to their shape and boundary structure.
We also compare our results with calculations based on the discrete dipole approximation as well as with an integral model for continuous particles, and analyse distinctions and similarities between the different approaches.
In particular, discrete metamaterials present themselves as a stringent platform for assessing continuous models developed for finite objects with sharp edges.
Overall, the reported results should be important for understanding mesoscopic systems with strongly interacting elements.
\end{abstract}

%%%%%%%%%%%%%%%%%%%%%%%%%%%%%%%%%%%%%%%%%%%%%%%%%

\maketitle

%%%%%%%%%%%%%%%%%%%%%%%%%%%%%%%%%%%%%%%%%%%%%%%%%%
%%%%%%%%%%%%%%%%%%%%%%%%%%%%%%%%%%%%%%%%%%%%%%%%%%

\section{Introduction}

Metamaterials, and more recently so-called metasurfaces, have been enjoying research attention for a good quarter of a century by now. Yet, not only many new challenges and suggestions still arise, but also some of the very fundamental questions regarding their theoretical description are not quite fully understood. To this end, various aspects of effective-medium theories for metamaterials description are particularly 
challenging \cite{Sim09,AgrGar09,Sil7,Mil10,Alu11}. 
On the one hand, metamaterials pose a great advantage for theoretical evaluation, since they are artificially made and thus offer direct control and exact knowledge over their internal structure,  
so that microscopic theory to describe metamaterial response can be built from first principles \cite{GorLapTre06}. 
Thus, for a number of practically relevant cases, effective-material models have been successfully developed, including for example 
artificial artificial magnetism with ring resonators \cite{GorLapSha02,BaeJelMar08}, 
artificial plasma with wire media \cite{BelMarMas03};
certain effects of disorder \cite{GorGreSha06}, noise \cite{SymSol11} and randomness \cite{AndLavPet16};
non-resonant wide-band diamagnetics \cite{LapKryBel13} and artificial chiral structures \cite{GorDmiEzh15}.

On the other hand, the very nature of metamaterials leads to numerous complications: a great variety of these are related to enhanced spatial dispersion \cite{Agranovich}, 
particularly relevant for metamaterials with extreme parameters \cite{GorRya98,SilBel8}
as well as for structures with one- \cite{GorGlyHur16,LapGor23},
two- \cite{BelMarMas03,ShaShvSir06},
and three-dimensional \cite{SilBaeJel09,GorVoyLap16} periodicities.

In relation to practice, effective-material description implies treating metamaterial structures as bulk media, often with very special edge effects introduced via transition layers \cite{VinMakRoz99,GadSuk00,VinDorMer10} 
and/or additional boundary conditions \cite{BelSim06,Yaghjian201070}. 
However, the core point that drives the difference between effective medium theory and observable properties of metamaterial samples is the finite size of any practically feasible structures. 
For conventional materials, even though surface effects and transition layers are not unheard of, there is no issues with attributing bulk parameters for evaluating their general properties, because the number of atoms in a sample is normally a figure of many orders of magnitude. 
For metamaterials, the effects of finite size arise through sample boundaries, and % for sufficiently small samples 
even additional boundary conditions may not be sufficient to reconcile effective-medium response with any problem that considers matching to the free space.
Interestingly, when it comes to discrete structures with clearly defined lattices, the problem actually becomes more severe, as we will further demonstrate in this article.

Indeed, it is already known that attempts to employ `bulk' theoretical models to predict the performance of manufactured metamaterial devices reveal significant discrepancies \cite{LapJelFre10,ShcPodBel14}.
For this reason, it is important to advance methods for reliable simulation of electromagnetic parameters of finite metamaterial samples, as well as to assess deviations of effective-medium approaches from actual properties of such mesoscopic systems.

At the same time, metamaterial structure can be deliberately arranged with a single unit cell (or single meta-atom) precision, resulting in arbitrarily perfect planes, straight edges, and sharp corners.  
This advantage provides an interesting alternative to address a long-standing problem of polarisability of small particles with sharp edges, most importantly, cubic particles. Starting from the seminal work of Fuchs \cite{Fuchs1975}, the spectrum of polarisability (or any related quantity) of a quasi-static cube has been represented as a superposition of six or more normal modes \cite{Langbein1976,Ruppin1996}. 
However, this multi-modal description is not fully consistent across literature.
This is not a problem for dielectric cubes, as the dependence of polarisability on the real susceptibility is a smooth function amenable to accurate rational interpolation \cite{Avelin2001,SihYlaJar04}. 
However, it becomes problematic for metamaterials with negative effective parameters \cite{WalKetSih08mm}, 
as well as for light scattering at plasmonic, especially silver, nanocubes \cite{Sosa2003,Zhou2008}. 
More recent articles show that peaks pattern evolves non-trivially with refining discretisation --- not only the peaks shift, but also new ones appear \cite{Klimov2014,Markkanen2017}. 
The problem originates from singular behaviour of fields near wedges or corners, which lead to nonphysical (infinite-energy) solutions for negative permittivity or permeability.
The above issues were recently settled by Helsing~\&~Perfekt~\cite{HelPer13}, who developed a stable surface-integral method which allows us to determine cube polarisability. 
We will assess the applicability of their method with the help of discrete metamaterial cubes.

%%%%%%%%%%%%%%%%%%%%%%%%%%%%%%%%%%%%%%%%%%%%%%% 
%%%%%%%%%%%%%%%%%%%%%%%%%%%%%%%%%%%%%%%%%%%%%%% 

\section{Methodological remarks on discrete metamaterials}

Without much loss of generality, we consider a prominent sub-topic within the metamaterial research: metamaterials for resonant artificial magnetic response. 
These can be implemented as an array of various kinds of (effectively) ring resonators \cite{MarJelFre11} 
which can be regarded as resonant linear contours in case for a deeply subwavelength regime.
Such resonators are excited by the magnetic field of an external electromagnetic wave, orthogonal to the ring plane, and the induced currents are inductively coupled in an array, contributing to the local magnetic filed inside the structure. 
The effective magnetic permeability for such metamaterials (bulk-medium approximation) can be obtained almost analytically for a given set of structure parameters (resonator impedance and lattice constants), by evaluating local fields and performing secondary macroscopic averaging \cite{GorLapSha02} under quasi-static approximation, taking the mutual interactions into account.
However, macroscopic averaging relies on the identity of all the unit cells, which excludes boundary effects and implies an effectively infinite structure.

For a finite size, instead, a complete set of Kirchhoff equations can be written, whereby inductive interaction between each resonator and all the other ones is explicitly taken into account. 
For any desired external magnetic field pattern, electromotive forces induced in every conductive ring can be obtained analytically, and the entire system then solved to find currents induced in each resonator \cite{LapJelMar10},
yielding good agreement to experiments.
%Further on, eventual magnetic field in every point can be calculated
%such a self-consistent calculation yields a nearly perfect match to experiments \cite{AlgFreLop11}. 
Using this approach, preliminary results have been reported \cite{LapJelMar12,LapMcPPou16}  
regarding the mismatch between effective-medium treatment and the role of boundary effects.
It was found that a typical unit cell design for isotropic magnetic response (implying three mutually orthogonal subsets of rings), while having complete translational symmetry in the bulk, leaves room for an ambiguity at the boundaries of the structure, and this ambiguity makes a huge difference to the observable response \cite{LapJelMar12}. 
Moreover, even spherical metamaterial samples do not behave quite like the effective-medium theory predicts. The convergence between the two is rather slow, as tested for samples with up to 20 thousand individual resonators \cite{LapMcPPou16}.

The above preliminary results left a number of questions to be answered through the analysis of substantially larger systems. However, direct computations based on the matrix inversion within the coupled impedance method \cite{LapJelMar10} 
cannot be employed in this case.
In this article, we circumvent this obstacle by adopting powerful methods of numerical analysis and linear algebra, taking into account a specific structure of matrices arising in periodic discrete systems. 
Indeed, regular spatial distribution of meta-atoms results in their mutual interaction being described by block-Toeplitz matrices, which can be stored with linear growing memory requirements, and multiplied by vectors with $\mathcal{O}(N\log N)$ asymptotic complexity \cite{GolubVanLoan}.
Replacement of the direct method with an iterative one for solution of linear algebraic equation systems, empowered with fast matrix-vector multiplications at each iteration, offers rigorous analysis of extremely large systems of fully coupled meta-atoms.

This approach is analogous to what is implemented within the \emph{discrete dipole approximation} (DDA) / \emph{method of moments} on regular meshes (MoM) \cite{LakhtakiaMoM1992,YurkinDDA2023},
widely used within analysis of linear electromagnetic scattering phenomena, %\cite{EvlReiEvl13,AndKuzLav15}
upon ideas presented already in 1970-s \cite{Bojarski1972}.

Note that, for the system of closely spaced rings that we consider, dipole approximation is not applicable directly because near-neighbour interaction of closely positioned rings is significantly different from that of two dipoles. 
To this end, the DDA approach (as a system of dipoles imitating the same resonance as predicted by the relevant effective medium theory \cite{GorLapSha02}) is going to deviate from the exact solution, particularly with low discretisation levels.
Nevertheless, it must be expected that in the limit of huge size, any discrete method should eventually converge to a continuous model of the bulk. 

Finally, we emphasise that our goal is to consider the effects of discrete structure and boundary ambiguity in their purity, so we consider an entirely quasi-static regime so that retardation effects \cite{ZhuSydSha09} and dimensional resonances \cite{ZhuShaSol05}, related to magnetoinductive waves \cite{ShaKalRin02jap}, do not come into play. 
Thus electric and magnetic fields are effectively decoupled and excitation can be treated as uniform external magnetic field harmonically varying in time.
Corresponding systems are practically feasible based on resonators with sufficiently low resonance frequency (see Sec.~\ref{sec:parameters}).
At the same time, quasi-static condition makes electrostatic DDA simulations equivalent to magnetostatic ones.

%%%%%%%%%%%%%%%%%%%%%%%%%%%%%%%%%%%%%%%%%%%%%%% 
%%%%%%%%%%%%%%%%%%%%%%%%%%%%%%%%%%%%%%%%%%%%%%% 

\section{System under study}

We are going to compare (i) actual response of discrete systems with strong mutual interaction, calculated precisely as described in Sec.~\ref{sec:dmm}, to (ii) continuous model (analytical or semi-analytical) based on the effective permeability corresponding to their unit cell (Sec.~\ref{sec:bulk_sim}), as well as to (iii) imitation of the same system with a DDA model (Sec.~\ref{sec:bulk_sim}).

We will analyse cubic particles as main object of interest, but also spherical particles where analytical theory is available. 
%Note that the system is entirely quasi-static regardless of the particle size.

For the role of an observable target function for this comparison we, most naturally, select magnetic polarisability of an entire finite-size sample \cite{JelKraCap17}, as a function of frequency in the relevant range above effective permeability resonance, where polarisability resonances are observed for cubes and spheres.

% % % % % % % % % % % % % % % % % % % % % % % % % % % % % % % % % % % % % % % % % % % % % % % % % % 

\subsection{System geometry: cubes}
\label{sec:cube-geom}

\begin{figure}[t]
	\centering
    \includegraphics[width=0.32\linewidth]{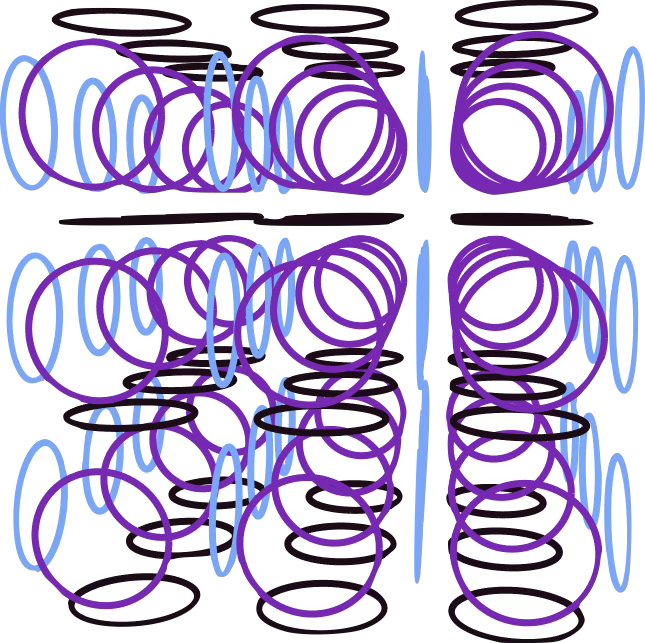}
    \includegraphics[width=0.32\linewidth]{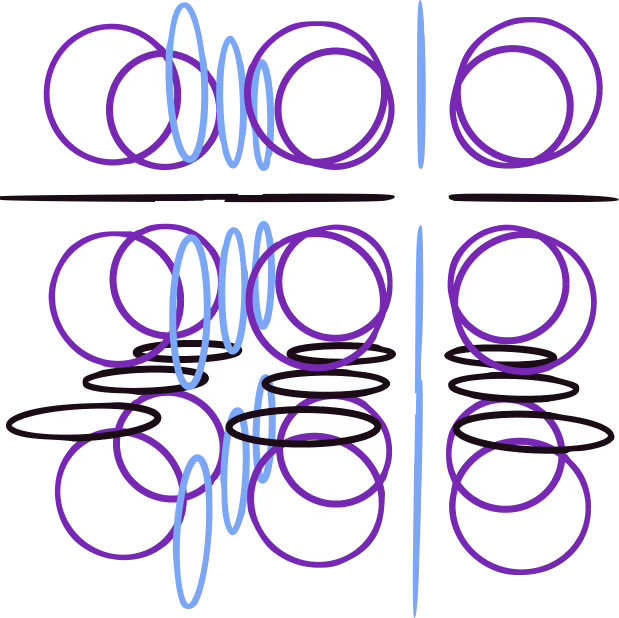}
    \includegraphics[width=0.32\linewidth]{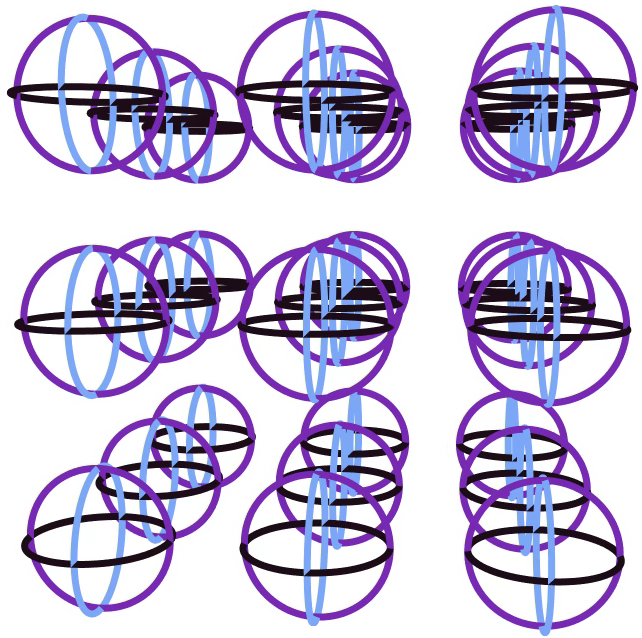}
    \caption{(colour online) Three types of discrete structures under consideration: smooth with boundaries terminated with ring planes (left), ragged with open boundaries (middle), and centred, with a symmetric unit cell (right).}
    \label{fig:structures}
\end{figure}

In order to achieve quasi-isotropic response, three otherwise identical but mutually orthogonal sub-lattices are used, each being a periodic array of circular resonators (rings). % parallel to one of the Cartesian coordinate planes.
From a practical point of view, the easiest way to construct such metamaterial is to place the rings as if they are located on the sides of a cube (left and middle pictures shown in Fig.~\ref{fig:structures}); such a corner of rings can form a periodic structure with perfect translational symmetry inside the structure.
This unit cell is not centro-symmetric itself, however for an infinite structure (as it is implicitly assumed for effective-medium treatment) the internal structure is still translationally symmetric.
At the boundaries of metamaterial however, an ambiguity with structure termination arises: if the entire sample is to be made symmetric, then either it should be terminated with planes of rings on each side (left picture of Fig.~\ref{fig:structures}), which we will call a `smooth' structure, or there should be no terminal ring planes on all sides (middle picture of Fig.~\ref{fig:structures}), which we will call a `ragged' structure.
Both versions have identical structure of a unit cell and are the same inside, and thereby they are subject to the same effective material parameters; the only difference arises at the boundaries.
Earlier research demonstrated strong difference between these two structures for relatively small cubic samples \cite{LapJelMar12}. 

In addition, we also consider cubes where the unit cell is itself symmetric \cite{BaeJelMar06}, as achieved by placing three orthogonal rings with a common centre; this will be called a `centred' structure (right picture of Fig.~\ref{fig:structures}). 
Note that this is not a convenient setup from a practical point of view, as the rings has to be shaped additionally so that they do not intersect and have no electric contact.
But from the effective-medium perspective, this is still the same structure as the two versions described earlier, as long as lattice constants are the same. This follows from cancellation of interactions of orthogonal rings in an infinite lattice due to the symmetry, i.e. each component of effective permeability is determined independently by the corresponding sub-lattice of rings. Conveniently, the centred structure has no ambiguity for the boundaries and constitutes a single geometric version. %, as shown in the right part of Fig.~(\ref{fig:structures}).

For each of the lattice type, cubes of various size will be considered to study the boundary effects (and their interplay with an elementary-cell configuration) as a function of system size.

% % % % % % % % % % % % % % % % % % % % % % % % % % % % % % % % % % % % % % % % % % % % % % % % % % 
% % % % % % % % % % % % % % % % % % % % % % % % % % % % % % % % % % % % % % % % % % % % % % % % % % 

\subsection{System geometry: spheres}
\label{sec:sph-geom}

%Likewise, slightly distinct 
Spherical samples can be produced from the three types of cubes described above, by truncating these cubes into a spherical shape as good as possible for a given size. 
Certainly, for small sizes, spheres shaped from a rectangular lattice are ragged with large steps all around, but with increasing size their surface becomes smoother. 
To this end, size effect for the spheres is mostly connected to imperfection of their shape as well as to the configuration of boundary rings at the surface, a small difference arising from using different source cubes.
In particular, truncation may seem to make the source cube models nearly undistinguishable, except for tiny patches of rings inherited from a smooth shape; yet the difference was known to be noticeable for at least up to 23 unit cells in diameter \cite{LapMcPPou16}.
We should also note that the centred configuration of the unit cell, albeit appearing more symmetric itself, is less advantageous for truncation, resulting in a more rugged shape for a given size.

An important difference to a cubic shape is that the total matrix of mutual impedances within a sphere would not be block-Toeplitz if only the actual rings are considered. In order to retain the advantageous matrix structure, removal of the unnecessary rings from the cubic source shape is technically achieved by formally retaining these within the interaction matrix however attributing infinite resistance to the `absent' rings, as explained in the next subsection. This is completely analogous to the concept of void dipoles/voxels used in the DDA \cite{YurkinDDA2023}.

% % % % % % % % % % % % % % % % % % % % % % % % % % % % % % % % % % % % % % % % % % % % % % % % % % 

\subsection{Technical parameters}
\label{sec:parameters}

As we are studying conceptual phenomena, specific parameters of the resonators are not particularly important. Still, we have used a realistic setup employed for metamaterial lens \cite{LapJelFre10}, 
with a ring radius $r = 0.49$\,cm and self-inductance $L = 13.5$\,nH, 
however assuming a much higher capacitance $C = 47$\,nF for deeply subwavelength operation, 
and resistance $R = 0.002$\,Ohm so that the quality factor of the resonators is about 270 (this will be important for some of the observed results).
Given the assumed quasi-static limit, such resonators are characterised with self-impedance function
$Z = - \mni \omega L + {\mni}/({\omega C}) + R$.
Then the individual resonance of a single ring thus occurs at $\omega_0 = 6.33$\,MHz. 
An array of these rings has a lattice constant $a = 15$\,mm, which is about 3000 times smaller than the free-space wavelength of 45--50\,m in the frequency range of interest, and the entire largest sample we considered is then 30 times smaller than the free-space wavelength. 
This said, note that our calculations here are quasi-static by essence, so this relationship to the wavelength does not play any role; we only mention these dimensional characteristics to indicate that making our system quasi-static is realistic with practically feasible parameters.
A detailed study of dynamic effects is certainly an important next step but that is retained for future research.
%Note also that maximal value of mutual inductance in this setup is still about 50 times smaller than the self-inductance, which we certainly would not call an excessively strong coupling.

For a more universal presentation of results, we will use relative frequency, normalised to $\omega_0$, or a relative shift measured in percents.
For reference, effective permeability has a single Lorentzian resonance at $\approx 0.987 \omega_0$ for the above parameters; real part equals to $-1$ at $\approx 1.0283 \omega_0$ and crosses zero at $\approx 1.0756 \omega_0$. As we will see further, the most interesting behaviour of the polarisability function happens between these reference frequencies, but not very close to them.

% % % % % % % % % % % % % % % % % % % % % % % % % % % % % % % % % % % % % % % % % % % % % % % % % % 
% % % % % % % % % % % % % % % % % % % % % % % % % % % % % % % % % % % % % % % % % % % % % % % % % % 

\section{Bulk-medium and discrete-dipole simulations}
\label{sec:bulk_sim}

With a given effective permeability function, based on metamaterial internal structure, we can evaluate its polarisability as if it were a bulk finite-size sample. 
For the cubic lattice considered in this article, relative effective permeability \cite{GorLapSha02} takes the form
\begin{equation}
\mu(\omega) = 
1 - \left(	 \dfrac{\mni a^3 Z(\omega) }{\omega \mu_0 \pi^2 r^4}
+
\dfrac{ a^3 \Sigma}{\pi^2 r^3}
+
\dfrac{1}{3}
%  }
 \right)^{-1} 
    \label{eq:eff_mi}
\end{equation}
where $\Sigma$ is a dimensionless parameter characterising mutual interaction, resulting from the summation over mutual impedances within a macroscopically large volume in the lattice \cite{GorLapSha02}. 
This functions features a single Lorentzian-type resonance.

For spherical samples, magnetic polarisability is well-known and is directly obtained analytically: 
\begin{equation}
\alpha_0 = 3 (\mu - 1) / (\mu + 2)
    \label{eq:spheralpha}
\end{equation}
featuring a single resonance at $\mu = -2$.

By contrast, cubic samples are much more complicated. As we have mentioned in the Introduction, Helsing~\&~Perfekt~\cite{HelPer13} have developed a stable surface-integral method to determine, first, the auxiliary function $\alpha^+$, which is the limit of cube polarisability when permeability approaches the negative real axis. Such limit is well defined, in contrast to the corresponding solutions for potentials or fields, and features two broad peaks (a continuous spectrum) instead of multiple discrete resonances. Moreover, the polarisability for any complex permeability can then be obtained through a stable integral representation:
\begin{equation}
    \alpha(z)=\frac{1}{\pi}\int_{x_1}^{x_2}{\mathrm{d}x\frac{\operatorname{Im}[\alpha^+(x)]}{x-z}}
    \label{eq:ref_polarisability}
\end{equation}
which follows from Eqs. (7), (13), (21), (25) of Ref.~\cite{HelPer13}. 
Here $z=(1+\mu)/(1-\mu)$ , while $x$ is the real integration variable related by the same transformation as $z$ to negative real $\mu$.
The integration boundaries $x_1\approx-0.695$ and $x_2=0.5$ are determined by the properties of the integrand, i.e.\ where $\operatorname{Im}[\alpha^+(x)] \neq 0$ (see Fig.~7 of Ref.~\cite{HelPer13}), and correspond to the continuous spectrum of the scattering operator. 
In terms of $\mu$, this corresponds to the interval $\mu\in[-5.55,-0.33]$.
%All the details about derivation and behaviour of function $\alpha^+(x)$ can be found in Ref.~\cite{HelPer13}.
We emphasise that Eq.~(\ref{eq:ref_polarisability}) is applicable for any $\mu$, but the most interesting effects are expected in the vicinity of the above continuous-spectrum interval.

Unfortunately, Ref.~\cite{HelPer13} is not much known in the community, partly on the premises that sharp boundaries never occur in practice. However, the practical consequence of Ref.~\cite{HelPer13} is that the response is very sensitive to both surface details and absorption, which is manifested, e.g., for rounded cubes \cite{HelPer13,Klimov2014,Tzarouchis2019}. Therefore, although the quasi-static limit of polarisability spectrum of Ref.~\cite{HelPer13} is not necessarily achievable in practice, no other universal limit exists. This makes a cube an excellent stringent test for metamaterials. 

We further denote Eq.~(\ref{eq:ref_polarisability}) as the Helsing-Perfekt solution, or simply HP-model, and implement it by applying the trapezoidal rule to dataset of 1195 values of $\operatorname{Im}[\alpha^+(x)]$, corresponding to Fig.7 of Ref.~\cite{HelPer13}. It was tested to provide accurate results for $\operatorname{Im}{\mu} \geq 0.003$.

Finally, the same metamaterial effective permeability is employed in the DDA simulations of quasi-static polarisability. For this purpose we utilise the open-source implementation ADDA \cite{Yurkin2011,ADDAgithub} v.1.5.0-alpha and obtain the polarisability from the amplitude scattering matrix at forward direction. We tune the simulation parameters to best reproduce the continuum results with respect to both accuracy and convergence of the iterative solver. Specifically, we use the IGT$_\text{SO}$ formulation of the DDA, which integrates the Green's tensor over the cubical voxel \cite{Smunev2015} using analytic approximations (exact in quasi-static limit) \cite{Yurkin_IGT_2023}. This approach resembles the MoM and involves the interaction term that significantly differs from that of point dipoles at short distances. However, both of them differ from the exact interaction of meta-atoms, so the DDA results can be considered as just another discrete approximation to the continuum one. 

Note that the DDA formulation is essentially dynamic so additional precautions were made to fulfil the quasi-static requirement, 
To make sure that we entirely exclude any retardation effects from DDA results, we have formally made all the geometric sizes 1000 times smaller while keeping the same resonant characteristics and the same effective permeability. 
In this way, even the largest DDA sample (256 dipoles per edge) is still deeply subwavelength. 

The convergence threshold of the iterative solver was set to $10^{-3}$, other simulation parameters are set to default values. For spheres, we additionally turned off the volume correction of the voxel lattice, which is used by default to ensure that that this volume exactly equals to that of the particle but is quasi-random with respect to grid refinement \cite{Yurkin2011}. However, the effect of this setting is much smaller than the ones discussed in the results (data not shown). The number of voxels per linear dimension (sphere diameter or cube edge) was increased to 256, which approximately corresponds to current feasibility when doing repeated simulations on a modern laptop.

% % % % % % % % % % % % % % % % % % % % % % % % % % % % % % % % % % % % % % % % % % % % % % % % % % 
% % % % % % % % % % % % % % % % % % % % % % % % % % % % % % % % % % % % % % % % % % % % % % % % % % 

\section{Numerical methods for discrete metamaterials}
\label{sec:dmm}

As we have mentioned already, for finite discrete systems a complete set of Kirchhoff equations can be written, whereby inductive interaction between each resonator and all the other ones is explicitly taken into account.
We also reiterate that for the purposes of this article we use quasi-static regime so that electric and magnetic fields are effectively decoupled and excitation can be considered as external magnetic field harmonically varying in time with the $\exp(\mni \omega t)$ multiplier, but uniform in space. 
% Thus the resonance frequency of the rings is such that the resonant free space wavelength is many orders of magnitude larger than the total size of the metamaterial samples considered; this can be technically achieved by using sufficiently large capacitors for the rings.

\subsection{Direct solution for discrete systems}

A system of coupled equations, which rigorously describes discrete metamaterials in the quasi-static limit, relies on the fact that the magnetic flux through a given ring number $m$ of a metamaterial sample is a superposition of fluxes from all rings in addition to any externally applied flux:
\begin{equation}
    \Phi_m = \sum_{n \neq m} L_{mn} I_n + \Phi_m^\text{ext}
    \label{eq:fulx_superposition}
\end{equation}
where the summation is performed over all the other rings using the corresponding mutual inductances $L_{mn}$.
With the resulting electromotive force, Ohm law yields a well-known system of coupled equations
\begin{equation}
    \sum_n \left[ Z_m\delta_{mn} + \mni\omega (1-\delta_{mn}) L_{mn} \right] I_n = -\mni\omega\Phi_m^\text{ext}
    \label{eq:coupled_current}
\end{equation}
whereby self-impedances $Z$ depends on frequency and is responsible for resonant behaviour, while the mutual impedances affect the overall characteristics but do not depend on frequency themselves.

Now, for the sake of numerical efficiency, we must reformulate these equations and utilise a complimentary set of equations for unknown complete electromotive forces in each ring:
\begin{equation}
    \sum_n \left[ \delta_{mn} + \mni\omega (1-\delta_{mn}) L_{mn} Y_n \right] \mathcal{E}_n = -\mni\omega\Phi_m^\text{ext}
    \label{eq:coupled_electromotive}
\end{equation}
The technical advantage of this equivalent representation is that in this way any absent rings can be implemented as virtual rings, formally contributing to the matrix of mutual impedances but having zero admittances, so that they do not contribute to  metamaterial electromagnetic response, but the block-Toeplitz matrix structure (see the next subsection) can be retained for an arbitrary sample shape. 

Solution to these equations yields the actual currents induced in each ring, and then the total magnetic polarisability of the entire sample can be obtained using the sum of all the individual magnetic moments.

% % % % % % % % % % % % % % % % % % % % % % % % % % % % % % % % % % % % % % % % % % % % % % % % % % 

\subsection{Numerical procedures for large systems}

A direct solution to system (\ref{eq:coupled_current}) requires a numerical row-reduction operation, which has the $\mathcal{O}(N^3)$ asymptotic numerical complexity in case of the direct inversion, and $\mathcal{O}(N^2)$ complexity in case of an iterative linear system solution, as well as $\mathcal{O}(N^2)$ memory usage in the both cases. This becomes rapidly impractical for large sizes.

To reduce the complexity we notice that the mutual inductances $L_{mn}$ are the same for any pair of rings with fixed orientations and relative shift of the centres along the three coordinate axes. Since we have three types of ring orientations, namely, with unit normal vectors to the ring planes directed along $X$, $Y$ or $Z$ axis, the set of all $L_{mn}$ can be split into 9 subsets $L_{mn}^{\alpha\beta}$ corresponding to magnetic interactions between different rings with fixed orientations (here $\alpha$ and $\beta$ stand for coordinate indices $x$, $y$, and $z$ corresponding to the ring normal direction). Note that many of $L_{mn}^{\alpha\beta}$ appear to be the same, and this symmetry is used to reduce the number of matrix elements to compute explicitly in simulations. 

Next, assuming first a metamaterial sample of a parallelepiped shape we notice, that each sub-matrix $\{ L_{mn}^{\alpha\beta} \}$ is a 3D-Toeplitz matrix due to the translational invariance of mutual inductances along each coordinate axis. In other words, expanding the ring index into a set of three indices enumerating the rings along the coordinate axes $m=(m_x,m_y,m_z)$ the matrix elements appear to depend only on the corresponding index difference
$$ L^{\alpha\beta}_{mn} \equiv L^{\alpha\beta}_{(m_x,m_y,m_z),(n_x,n_y,n_z)} = L^{\alpha\beta}_{(m_x-n_x), (m_y-n_y), (m_z-n_z)} $$
It is well-known that any Toeplitz matrix can be expanded into a corresponding circulant matrix \cite{GolubVanLoan}, which is a matrix having each row obtained from a previous row by a cyclic shift by one element. A product of a circulant matrix by a vector is a discrete convolution product, which can be computed using the fast Fourier transform (FFT) very efficiently with $\mathcal{O}(N\log N)$ asymptotic numerical complexity.

Turning back to the equation system (\ref{eq:coupled_electromotive}) we write it in the matrix-vector form
\begin{equation}
    (\mathbb{I} + \mathrm{L}\mathrm{Y}) \bm{\mathcal{E}} = \bm{\phi}
    \label{eq:system_mv}
\end{equation}
where the diagonal matrix $\mathrm{Y} = \mathrm{diag}\{\mni\omega Y_m\}$, unknown vector of ring electromotive forces $\bm{\mathcal{E}} = \{\mathcal{E}_m\}$, the right-hand part excitation vector $\bm{\phi} = \{\mni\omega\Phi^\text{ext}_m\}$, and $\mathrm{L}$ is the $3\times3$-block 3D-Toeplitz matrix
\begin{equation}
    \mathrm{L} = \left( \begin{array}{ccc}
        L^{xx} & L^{xy} & L^{zz}  \\
        L^{yx} & L^{yy} & L^{yz}  \\
        L^{zx} & L^{zy} & L^{zz}
    \end{array} \right)
\end{equation}
Therefore, when solving the system for the unknown vector ${\mathcal{E}}$ with an iterative method like the BiCG or the {GMRes} \cite{Saad2003}, matrix-vector multiplications at each iteration are performed with $\mathcal{O}(N\log N)$ complexity. This makes the overall method extremely efficient, enabling analysis of extremely large arrays of meta-atoms even on a laptop.

Further on, this method can be extended to arbitrarily shaped samples. Indeed, diagonal elements of matrix $\mathrm{Y}$ can take any values without affecting the overall numerical complexity of computations. Thus, by assigning zero values of admittances for selected ring positions, such rings can be excluded from the original parallelepiped shape, which effectively produces an object of an arbitrary shape cut out from the originally rectangular block fully encompassing the desired object. Albeit matrix operations run for a full-size matrix in such cases, this procedure is nevertheless highly advantageous compared to a direct calculation made without the above Toeplitz-form optimisation.

% % % % % % % % % % % % % % % % % % % % % % % % % % % % % % % % % % % % % % % % % % % % % % % % % % 
% % % % % % % % % % % % % % % % % % % % % % % % % % % % % % % % % % % % % % % % % % % % % % % % % % 

% % % % % % % % % % % % % % % % % % % % % % % % % % % % % % % % % % % % % % % % % % % % % % % % % % 
% % % % % % % % % % % % % % % % % % % % % % % % % % % % % % % % % % % % % % % % % % % % % % % % % % 

% % % % % % % % % % % % % % % % % % % % % % % % % % % % % % % % % % % % % % % % % % % % % % % % % % 
% % % % % % % % % % % % % % % % % % % % % % % % % % % % % % % % % % % % % % % % % % % % % % % % % % 

\section{Results: cubic metamaterial samples}

Armed with the new protocol to analyse much larger structures than previously, we proceed with calculation of magnetic polarisability of the cubes containing up to 100 unit cells along their sides.
We calculate magnetic polarisability of such cubic samples in response to magnetic field along the axis of the rings of one of the sublattices, in other words, perpendicular to cube sides in one of the directions.  

\begin{figure*}
    \centering
    \includegraphics[width=0.99\textwidth]{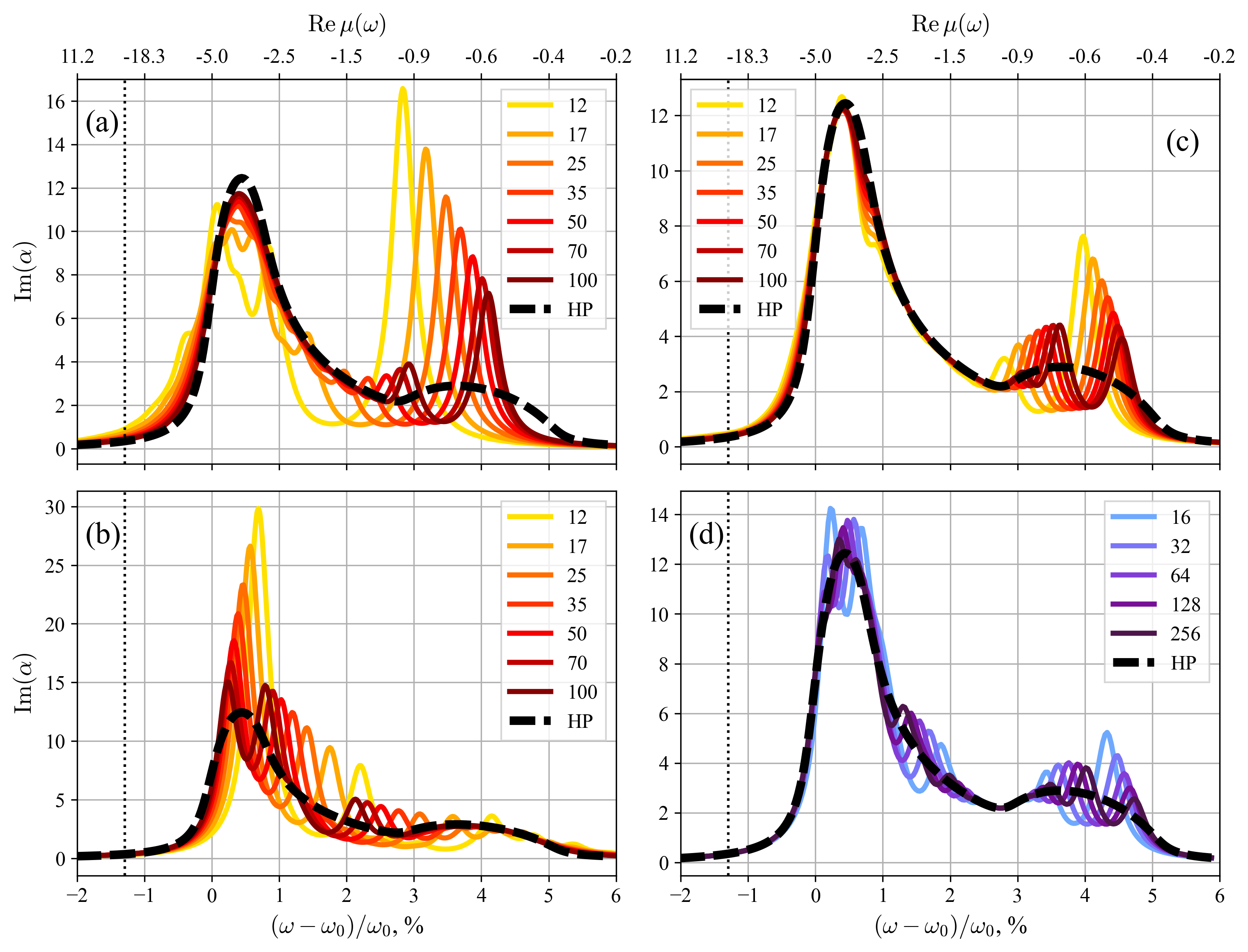}
    \caption{(colour online) Frequency dependence of the imaginary part of polarisability for discrete metamaterial cubic sample with increasing number of meta-atoms along the cube side from 12 to 100, for 
    (a) smooth; (b) ragged; (c) centred configurations, as well as (d) for the DDA model with increasing discretisation along the cube side in the range from 8 to 256 lattice constants.
    Resonance frequency of the corresponding effective permeability is marked with vertical dotted lines. 
    Note that the horizontal axes are the same in all the plots, showing frequencies at the bottom and the corresponding $\operatorname{Re}{\mu}$ at the top; however vertical scales are all different. The HP curve shows the same data across all the plots.}
    \label{fig:spectra_discrete}
\end{figure*}

For clarity, we remind that the problem is considered to be sufficiently subwavelength so that the electric size of the cube is not playing any role yet: we are dealing with entirely quasi-static behaviour.
We compare three types of internal structure described above: 
smooth, ragged and centred, see Fig.~\ref{fig:spectra_discrete}\,(a)--(c). 
We use relative frequency shift as a horizontal axis for this and further similar figures, and provide additionally a top axis showing $\operatorname{Re}{\mu}$ corresponding to that frequency.
Note that $\operatorname{Re}{\mu}$ rapidly changes from 22 to -20 in the vicinity of the resonance frequency, but the polarisability is quite small and shows no resonances in this region. 
The size of the cubes is indicated in terms of the number of unit cells along each edge.
It is important to note that the smooth and ragged versions have identical internal structure and the only difference appears at the surface of a sample; however the centred version has a different internal structure. 
At the same time, from the perspective of effective-medium treatment, all the structures correspond to the same effective permeability.

For each specific structure type, there is a certain trend for convergence within its size series, but even so we would not be confident predicting eventual spectral shape in the limit of infinite size, in spite of the huge size of largest systems (structures with 100 unit cells per side feature about 3 million individual resonators).
Between the types, however, we observe rather different trends, which can be more conveniently described in terms of comparison with the continuous HP model (the same thick black dash in all the plots).

The smooth structure, Fig.~\ref{fig:spectra_discrete}\,(a), exhibits rather irregular spectral shapes for smaller sizes but for larger sizes tends to reproduce the main peak of HP spectrum quite well.
At the same time, in place of the minor broad HP peak, smooth cubes show a much stronger and sharper peak, which decreases somewhat and blue-shifts with size, but remains quite pronounced; an additional secondary peak further appears between the two, in the frequency range corresponding to $\operatorname{Re}{\mu}\approx-1$. 
Similarly, the centred structure, Fig.~\ref{fig:spectra_discrete}\,(c), reproduces the main peak of the HP spectrum even better and increasingly well as the size grows, however also shows two stronger resonances in place of the broad minor HP-peak; contrary to the main peak which is well-settled in frequency already for the smaller sizes, the additional peaks show a steady blue shift.
To this end, spectra for the smooth and centred configuration appear to be qualitatively similar for large size.
In contrast, ragged cubes, Fig.~\ref{fig:spectra_discrete}\,(b), show two adjacent narrow peaks in place of the main HP resonance, but at the same time reproduce the minor broad part of the HP spectrum very well for large sizes.

So far, one might rush for a conclusion that, albeit all the structure types result in qualitatively different spectra, smooth type appears more similar to the centred type, while ragged type is the odd one.

We would like to emphasise once again that there is no difference in the bulk between smooth and ragged structures, so the very essential difference in response is entirely due to surface effects, however small the fraction taken by surface resonators may seem compared to those in the bulk of the structure.  
This finding highlights the major importance of surface effects and suggests that any modifications at the surface, that is, changes made to a very small part of resonators, would have a major effect over the properties of the entire sample.

Compared next to the DDA modelling, the latter shows a somewhat different convergence trend in Fig.~\ref{fig:spectra_discrete}\,(d). 
Contrary to the exact discrete models, its convergence towards the HP-model is approximately uniform across the entire frequency range: the DDA spectra deviate from the HP spectrum in the form of some ``oscillations'' of the DDA spectral curves relative to the HP curve.
More specifically, the DDA calculations approach the main peak of the HP-model better than ragged discrete cubes, but much worse than smooth and centred ones. 
At the same time, DDA calculation approaches the minor broad peak area of the HP-model better than the smooth and centred types, but much worse than ragged type. 
Still, this cannot be called a uniform convergence in the mathematical sense due to the shifting of all the peaks. 

Conceptual difference between the DDA and the full-model calculations, is the way mutual interaction of nearby resonators is taken into account. Point-dipole formulation of the DDA may be interpreted as assuming the rings to have the same individual resonance properties but having a negligibly small radius, compared to the lattice constants. The employed IGT formulation additionally smears the polarisability over a single meta-atom. The change of the DDA formulations is known to affect the eigenspectrum of the interaction matrix \cite{Smunev2015}, which mimics the polarisability spectrum for sufficiently small absorption. Limited point-dipole simulations for up to 32 dipoles along the cube edge suggest that the individual peaks shift, but overall trend (in comparison with the HP-model) remains the same (data not shown). While the DDA results do not match any discrete cubes, the difference is the same or even smaller than between different structures of the latter, suggesting that the surface details are at least equally important as the details of close-neighbour interaction. Apparently, none of the boundary types theoretically correspond to the DDA, but the latter seems to be not directly suitable for exact simulation of any of the considered cubical metamaterial samples.

\begin{figure}[t]
    \centering
    \includegraphics[width=0.49\textwidth]{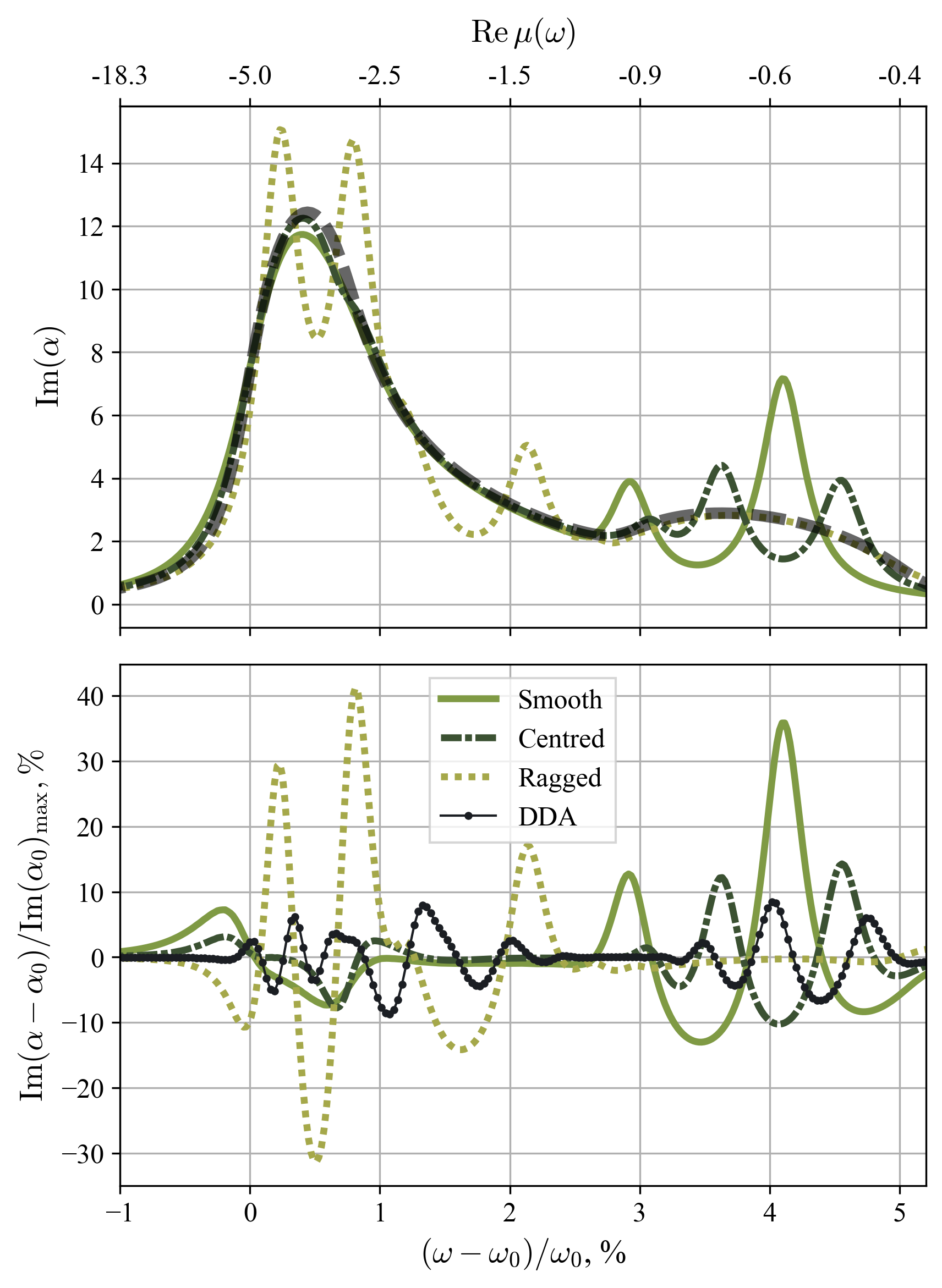}
    \caption{(colour online) Comparison of the frequency dependence of the imaginary part of the cube polarisability calculated for the maximum size (100 unit cells per edge) for the three possible structure types:
    actual values (top) and normalised difference from the HP-model (bottom). 
    The bottom plot also includes the difference between the HP-model and DDA result of the finest discretisation (256 voxels per cube edge).}
    \label{fig:kub-compare}
\end{figure}

To summarise the difference between the different approaches and the HP-model, we plot together the largest structures (100 unit cells per edge) of all the three types in Fig.~\ref{fig:kub-compare}\,(top). Next to that, Fig.~\ref{fig:kub-compare}\,(bottom) shows the difference for each type of discrete metamaterial as well as for the finest DDA discretisation. For clarity, the difference shown is normalised to the maximum value of the HP-polarisability.

Interestingly, when compared to the earlier study \cite{LapJelMar12} with up 15 cells per cube edge, the new findings for 20 to 100 size range lead to somewhat opposite conclusions.
Namely, the year 2012 impression was that the ragged structure appears to be more similar to continuous samples.
While the present calculations reproduce those earlier data for small sizes, larger sizes make the smooth version closer to the certainly more symmetric centred structure, and together they reproduce the main peak of the HP-model consistently, in contrast to the ragged structure.

Further insight into the difference between various structure types is provided with the distribution of currents induced in the individual rings within the structure.
Indeed, our calculation method explicitly resolves currents induced in all the rings of the structure and in theory this even permits to calculate local (meta-microscopic) fields within the entire structure, however this is an extremely memory-consuming process for large structures so only current magnitudes are calculated at present. 
Even so, we are not able to present the current distributions for the largest cubes and only show these for size-50 structures; given the similarity of the polarisability spectra, we do not expect any qualitative difference.

Animated files in the supplementary materials illustrate how the distribution of currents, excited in the individual resonators across the structure, changes with frequency. 
For each structure type with a size of 50 unit cells per edge, 
we depict four planar cross-sections showing currents induced in those rings perpendicular to applied field:
two planes of resonators parallel to the applied field ($xz$-planes),
and two planes perpendicular to the applied field ($xy$-planes),
for each orientation  
one plane at the face of a cube (1st layer), 
the other plane in the middle of a cube (25th layer).
Please see the supplementary materials with the files 
\texttt{cube-S.mp4} for smooth, 
\texttt{cube-R.mp4} for ragged and 
\texttt{cube-C.mp4} for centred structures.
All the animations are supplemented with the corresponding polarisability spectrum showing the running frequency with a moving vertical line.
Note that for better visibility the currents are dynamically normalised to the maximum at each given frequency, however the absolute magnitude is reflected with the help of dynamically changing magnitude colour bars. 

Inspection of the frequency dependence of the distribution of currents reveals essential differences between various structure types, as well as certain common trends. 
Overall, smooth structure provides a more rich and sophisticated pattern of spatial distribution, particularly at the sides of the cube. 
Ragged and centred structures provide more clear patterns, mostly similar to each other at the comparable frequencies. 

Around the main resonance of the HP model, which is also the main resonance for smooth and centred structures, cube corners are strongly excited; with ragged and centred structures, everything else is almost negligible, and for smooth structures, there are some further excitations on the faces perpendicular to the applied field, but the corners are still dominant. 
We should specifically note that in spite of the fact that ragged structure shows two adjacent resonances in place of the main HP resonance, current patterns are very similar for both the resonances and even for a local minimum between them.
To this end, we can say that the main resonance is associated with cube corners for all the structure types, albeit with very noticeable difference between smooth type and ragged or centred types; while the latter two are more similar to each other.

In the area of broad minor HP resonance, smooth structures have a single strong resonance where excitation is predominantly along the edges but not the corners. Similarly, centred structures are also mostly excited along the edges. However, ragged structures have no clear resonances in this spectral range, while the patterns of currents change a lot with frequency and typically show multiple hot spots on all the sides.

The overall trend is that the resonances evolve from cube vertices to edges and then faces with increasing frequency (or decreasing $|\operatorname{Re}{\mu}|$), which agrees with electrostatic simulations for cubes \cite{Klimov2014,Markkanen2017,Tzarouchis2019}.
However, for the discrete metamaterials considered here, a more detailed study of internal distribution is worth further attention and we plan to do so in the future.

Reiterating on the above observations, in terms of the polarisability spectra, centred and smooth structures appear similar and reproduce the main HP peak very well, whereas ragged structures reproduce the broad minor HP peak instead. 
However, in terms of the distribution of currents inside the cubes, centred and ragged structures turn our to be similar (particularly in the range main HP peak in spite of the different spectra), whereas smooth structures are much more different.

Detailed investigation of the role of boundaries on the distribution of excitations in relation to the spectra is the subject of future research. 
At this stage, we can only comment that the smooth structure includes resonators which are placed in the environment much different from the bulk, and so the boundary surfaces are likely to support ``their own'' excitation patterns, whereas in ragged structures such excitations are less pronounced, and observed at different frequencies. 
As to the centred configuration, the role ob boundary layers must be the smallest as there are no orthogonally oriented rings within the nearest neighbours, which is supported with the analysis of currents inside the structure.

Finally, we would like to discuss some implications of our results for a broader context of polarisability of cubic particles.
The results of Ref.~\cite{HelPer13} address a cube with sharp edges and corners, a situation which cannot be realistically achieved, for example, in plasmonic particles, due to their natural rounding.
With metamaterials, we can emulate sharp corners with the precision of a single unit cell, and in this way our structures are well-positioned to test the applicability of HP-model for sharp cubes.
To this end, thanks to a quasi-static consideration, increasing size of a cube may be also interpreted as increasing sharpness (as decreasing the ratio of characteristic scale of rounding to the particle size), particularly so for large sizes where collective effects in the bulk are well presented.
Indeed, in spite of certain specific deviations at some frequencies, our calculations across various structure types generally converge to the HP model.

At the same time, further data presented in Ref.~\cite{HelPer13} for rounding of other particles (2D super-ellipses) reveal that even for extremely small rounding scales some resonances are not settled.
Furthermore, the above issue is also strongly dependent on dissipation.
In our case, $\operatorname{Im} z \approx 0.045$ is sufficiently low (in comparison with the distance between discrete resonances of the rounded or discretised cubes~\cite{HelPer13,Klimov2014}) to reveal these interesting effects.
But for large dissipation, such as e.g.\ for gold cubes, DDA is known to converge well with refining discretisation~\cite{Yurkin2010}.
Thus, the role of boundary effects is likely to be suppressed with increasing absorption \cite{KinThiPar83,CheCarKoi16}, 
however even a huge size increase for our structures, such as 1000 times in each direction --- which is neither plausible for calculations nor  practically feasible --- may still not necessarily lead to a complete convergence of all our data to the HP curve.

Overall, the interplay between dissipation, rounding and absolute number of resonators is a complex problem which we plan to address in detail in a separate study.  

% % % % % % % % % % % % % % % % % % % % % % % % % % % % % % % % % % % % % % % % % % % % % % % % % % 
% % % % % % % % % % % % % % % % % % % % % % % % % % % % % % % % % % % % % % % % % % % % % % % % % % 

% % % % % % % % % % % % % % % % % % % % % % % % % % % % % % % % % % % % % % % % % % % % % % % % % % 
% % % % % % % % % % % % % % % % % % % % % % % % % % % % % % % % % % % % % % % % % % % % % % % % % % 

\section{Results: spherical metamaterial samples}

The reason to analyse large spherical samples in this paper is two-fold: to obtain further confirmation for the convergence observed earlier with smaller spheres \cite{LapMcPPou16}, and at the same time to verify our numerical procedures in comparison to the analytical solution. 

Given the known effective permeability \eqref{eq:eff_mi} for the considered metamaterial structure, polarisability of a homogeneous spherical particle $\alpha_0$, see Eg.~\eqref{eq:spheralpha}, is analytically obtained. 
As discussed in Sec.~\ref{sec:sph-geom}, there is a mismatch between the rectangular shape of metamaterial unit cells and a spherical shape of the sample (staircase effect). 
Earlier research on relatively small samples up to 20 thousand meta-atoms has indicated that spherical metamaterial samples demonstrate a reasonable, but slow convergence towards the analytical prediction~\cite{LapMcPPou16}.
We have now performed calculations for spherical samples with up to 100 unit cells per diameter, which corresponds to over 1 million individual resonators.

\begin{figure*}
    \centering
    \includegraphics[width=0.99\textwidth]{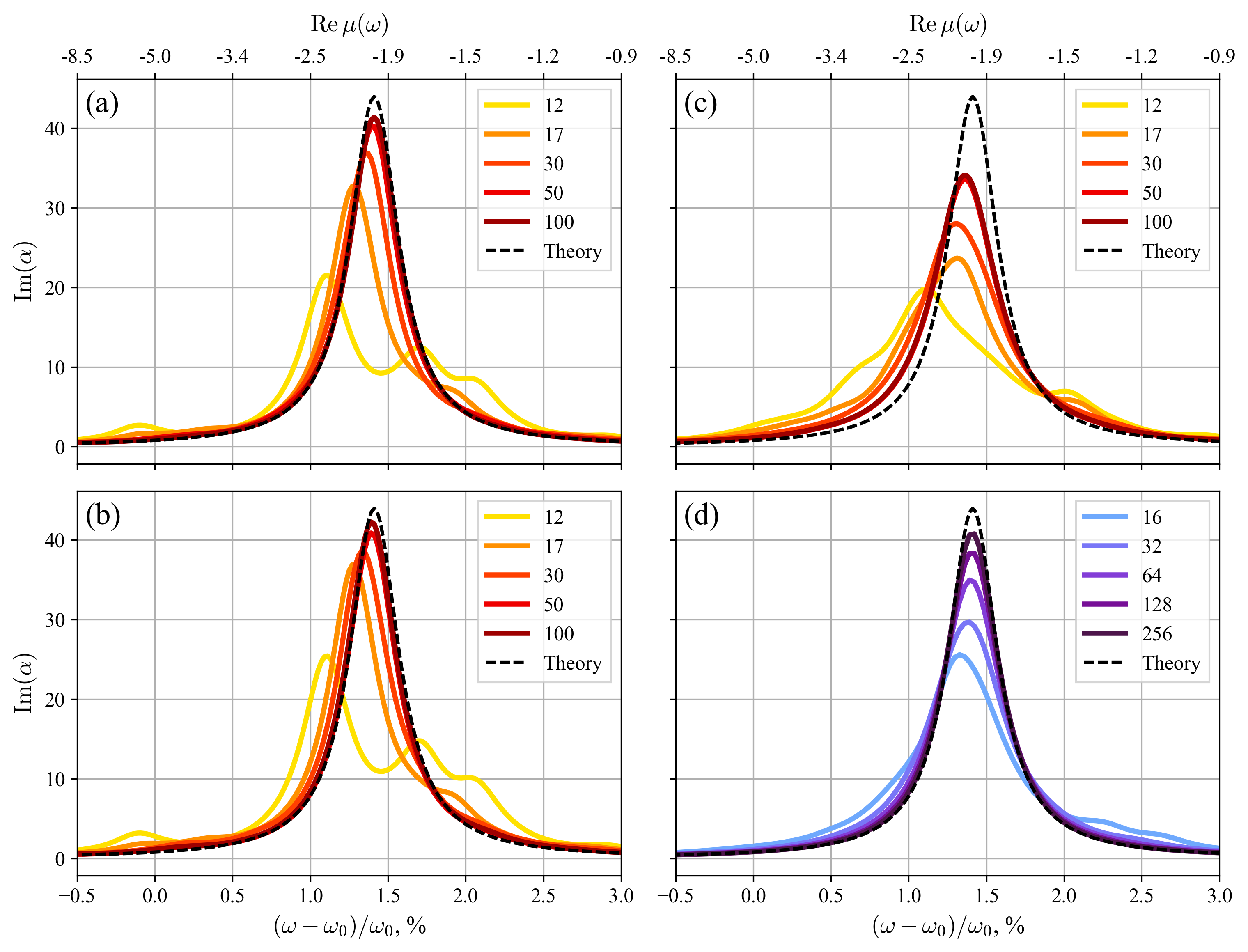}
    \caption{(colour online) Frequency dependence of the imaginary part of polarisability for discrete metamaterial spherical samples with increasing number of meta-atoms along the diameter, from 12 to 100: 
    (a) smooth; (b) ragged; (c) centred configurations, as well as (d) for DDA model with increasing discretisation along the diameter in the range from 8 to 256 lattice constants.
    Resonance frequency of the corresponding effective permeability is marked with dotted lines.
    The horizontal axes are the same in all the plots, showing frequencies at the bottom and the corresponding $\operatorname{Re}{\mu}$ at the top; and vertical scale is the same in all the plots.}
    \label{fig:sph-spect}
\end{figure*}

We compare the same three types of internal structure described above: smooth in Fig.~\ref{fig:sph-spect}~(a), ragged in Fig.~\ref{fig:sph-spect}~(b) and centred in Fig.~\ref{fig:sph-spect}~(c), and present polarisability spectra for various sizes.
Overall, smooth and ragged versions appear to be more similar whereas centred versions are more distinct from them.
Indeed, for the centred structures the staircase effect is stronger (due to a full unit cell size stair-steps) as compared to the other types.
The smooth and ragged versions have identical internal structure and the only difference appears at the surface of a sample, so when the original cube is truncated to a spherical shape, its surface is a kind of mixture between smooth and ragged scenarios. 
Nevertheless, some minor difference is still visible in the polarisability spectra. 

The DDA results for spheres, presented in Fig.~\ref{fig:sph-spect}~(d), support the same trend. While the sphere is known to be a challenging shape for the DDA, both for plasmonic nanoparticles \cite{Yurkin2010} and for morphologically-dependent resonances in larger dielectric ones \cite{Zhu2019}, reliable results can still be obtained with fine enough discretisation. The principal difference with cubic particles is that the spectrum of the polarisability for particles with smooth shapes is a discrete one (for sphere, a single point), to which the spectrum of discrete operator (for any of the employed methods) converges in a regular manner.

Further insight is provided by the animated distribution of currents, excited in the individual resonators across the structure, depending on  frequency.
For each structure type with a size of 50 unit cells per diameter, 
we present two planar cross-sections (through the centre of the sphere) 
showing currents induced in those rings perpendicular to applied field:
a plane of resonators parallel to the applied field ($xz$-plane),
and a plane perpendicular to the applied field ($xy$-plane).
Please see the supplementary materials with the files 
\texttt{sph-S.mp4} for smooth, 
\texttt{sph-R.mp4} for ragged and 
\texttt{sph-C.mp4} for centred structures.
All the animations are supported additionally with the corresponding polarisability spectrum showing the running frequency with a moving vertical line.
Note that for better visibility the currents are dynamically normalised to the maximum at each given frequency, however the absolute magnitude is reflected with the help of dynamically changing magnitude colour bars. 

In particular, these animations show that for all the cases distribution of currents is not uniform, in contrast to a uniform field inside a homogeneous sphere, theoretically expected.
Moreover, as frequency varies, the currents are individually changing in a very complicated way.
Nevertheless, it turns out that on average all these internal inhomogeneities are mutually compensated so that the overall macroscopic response of discrete spherical samples agrees increasingly well to the analytical result for a homogeneous sphere.

Finally, we note that observing the expected size dependence and evolution of spectra constitutes an additional verification of the enhanced computational procedures developed in this article.

% % % % % % % % % % % % % % % % % % % % % % % % % % % % % % % % % % % % % % % % % % % % % % % % % % 
% % % % % % % % % % % % % % % % % % % % % % % % % % % % % % % % % % % % % % % % % % % % % % % % % % 

\section{Conclusions}

We studied electromagnetic properties of volumetric metamaterials containing a huge amount of individual identical resonators (up to over 1 million) in a closely spaced periodic lattice.
At this stage, we only consider a strongly subwavelength problem so that quasi-static calculations can be used.
This assumption, albeit not very typical for metamaterials, is fairly consistent with natural structures, as the size of an atomic cluster is typically much smaller than the wavelength at which atoms resonate.  
We have considered resonators in the form of capacitively-loaded rings and their mutual interaction can be calculated exactly via mutual inductance, and no approximations are made with this respect, so that mutual inductances between absolutely all the resonators in the lattice are calculated and used to solve a full electromagnetic problem.
Note that the interaction of such resonators at close distances is quite different to dipolar interaction.
Such calculation involves huge matrices for mutual interaction and the corresponding linear system of Kirchhoff equations cannot be solved directly. 
We have developed a numerical procedure for solving the system based on the fact the the matrix of mutual admittances can be written in a block-Toeplitz form and then numerical methods for faster solutions and smaller memory requirements could be employed.

We have compared the actual response of finite-size metamaterial samples, calculated exactly, to the properties of continuous samples assigned with the effective permeability corresponding to their structure, as well as to the DDA simulation of the latter. For the metamaterial lattice that we consider, the same structure in the bulk can be symmetrically terminated at the boundaries in different ways, resulting in samples which correspond to the same effective permeability but different boundary conditions. 
Specifically, the subject of comparison is the overall magnetic polarisability of a finite-size object (a sphere or a cube), which in total is much smaller than the wavelength.

For spheres, we found a good convergence, so that with increasing size the polarisability calculated exactly for discrete structures, approaches analytical prediction for a homogeneous sphere with the corresponding effective permeability. 
Note, however, that the convergence is rather slow due to the shape mismatches when a sphere is assembled with a rectangular lattice. 

For cubes, however, different structures and approaches produce remarkably distinct results. 
For a homogeneous cube, the HP model predicts a major peak in the frequency ranges corresponding roughly to $\operatorname{Re}{\mu}\in[-6,-2]$, and a much smaller and broader peak in the range of $\operatorname{Re}{\mu}\in[-0.9,-0.4]$.
Discrete cubes behave quite differently depending on the boundary type, but none of structures reproduces the entire spectrum.
Smooth and centred types tend to reproduce the major HP-peak increasingly well with size, but show additional resonances in place of the minor flat HP-peak. 
Ragged structure, on the contrary, reproduces the minor peak very well however consistently shows two sharper peaks in place of the single major HP-peak; further evolution of these spectra remains inaccessible even with the efficient numerical procedures developed in this article. 
Then, DDA results converge to the the HP model more uniformly, showing comparable oscillations across both the peaks, but with the best convergence in the vicinity of $\mu=-1$.
The major remaining questions are to determine the threshold values of absorption for various non-smooth shapes and to develop a theoretical model for the dependence of the discrete operator spectrum on the boundary structure of metamaterials.

Importantly, our results generally support HP model as a reliable reference for a perfect homogeneous quasi-static cube. 
The eventual difference is due to the structure of system boundaries and thereby to the extent to which a precise discrete structure imitates a continuous object with sharp edges.
Further insight into these effects requires a systematic study of absorption effects aw well as boundary modifications, which is beyond the scope of the present paper.

On a positive side, our findings show that the role of the relatively few edge and corner resonators is huge in the overall electromagnetic response, thus paving a way towards efficient control of metamaterial properties without the need to modify the entire structure.

% % % % % % % % % % % % % % % % % % % % % % % % % % % % % % % % % % % % % % % % % % % % % % % % % % 
% % % % % % % % % % % % % % % % % % % % % % % % % % % % % % % % % % % % % % % % % % % % % % % % % % 

\section{Acknowledgement}
We thank Johan Helsing for providing the raw data for Fig.\,7 of Ref.~\cite{HelPer13}.
M.Y. (DDA simulations) acknowledges support of the Normandy Region (project RADDAERO).
A.M. and A.S. (development of the numerical method, simulation and analysis of the discrete metamaterial samples) acknowledge support by the Russian Science Foundation (grant no.~22-11-00153-$\Pi$). % П

% % % % % % % % % % % % % % % % % % % % % % % % % % % % % % % % % % % % % % % % % % % % % % % % % % 
% % % % % % % % % % % % % % % % % % % % % % % % % % % % % % % % % % % % % % % % % % % % % % % % % % 

\bibliographystyle{apsrev4-2}
\bibliography{kub_refs} 

\end{document}